\documentclass[aps,prd,amsmath,floats,floatfix,twocolumn,superscriptaddress,
  nofootinbib,showpacs]{revtex4} 
\usepackage{graphicx} 
\usepackage{bm}

\usepackage{amsmath}
\usepackage{amssymb}
\usepackage{psfrag}

\newcommand{\Cornell}{\affiliation{Center for Radiophysics and Space
    Research, Cornell University, Ithaca, New York, 14853}}

\begin{document}
\vspace{-2.5cm} 

\title{ Ineffectiveness of Pad\'e resummation techniques in post-Newtonian approximations}
\author{Abdul H. Mrou\'e} \Cornell
\author{Lawrence E. Kidder} \Cornell
\author{Saul A. Teukolsky} \Cornell

\date{\today}

\begin{abstract}
 We test the resummation techniques used in developing Pad\'e and Effective One
 Body (EOB) waveforms for gravitational wave detection. Convergence tests
 show that Pad\'e approximants of the gravitational wave energy flux do not 
accelerate the convergence of the standard Taylor approximants even in the test
 mass limit, and there is no reason why Pad\'e transformations should help in 
estimating parameters better in data analysis. Moreover, adding a pole 
to the flux seems unnecessary in the construction of these Pad\'e-approximated 
 flux formulas. Pad\'e approximants may be useful in suggesting the form of 
fitting formulas. 
We compare a 15-orbit numerical waveform of the Caltech-Cornell 
group to the suggested Pad\'e waveforms of Damour et al. in the equal mass, 
nonspinning quasi-circular case. The comparison 
suggests that the Pad\'e waveforms do not agree better with the numerical 
waveform than the standard Taylor based waveforms. 
Based on this result, we design a simple EOB model by modifiying the ET EOB model of 
Buonanno et al., using the Taylor
series of the flux with an unknown parameter at the fourth post-Newtonian  
order that we fit for. 
This simple EOB model generates a waveform having a phase difference 
of only 0.002 radians with the numerical waveform, much smaller than 0.04 radians  the 
phase uncertainty in the numerical data itself. An EOB Hamiltonian can make 
use of a Pad\'e transformation in its construction, but this is the only  
place Pad\'e transformations seem useful. 
\end{abstract}

% 04.25.D- Numerical relativity 
% 04.25.dg Numerical studies of black holes and black-hole binaries 
% 04.25.Nx Post-Newtonian approximation; perturbation theory; related approximations 
% 04.30.-w Gravitational waves (see also 04.80.Nn Gravitational wave detectors and experiments)
% 04.30.Db Wave generation and sources 
% 02.70.Hm Spectral methods 

\maketitle

%\tableofcontents

%#########################
\section{Introduction}
%#########################

Even though general relativity was developed at the beginning of the twentieth 
century,
no analytical solution is known for the two-body problem. Until recently, 
attempts to find a numerical solution failed because of the complexity of the
 mathematical equations and the instabilities inherent in the analytical
formulations being used. In the past few years, breakthroughs in numerical 
relativity~\cite{Pretorius2005a,Pretorius2006,Campanelli2006a,Baker2006a} 
allowed a system of two inspiraling black holes to be evolved through merger 
and the ringdown of the remnant black hole~\cite{Campanelli-Lousto-Zlochower:2006,Herrmann2007b,Diener2006,Scheel2006,Sperhake2006,Bruegmann2006,Marronetti2007,Etienne2007, Szilagyi2007}. 

Studying the late dynamical evolution of these inspiraling compact binaries is 
important because they are among the most promising source of gravitational
 waves for the network of laser interferometric detectors such as LIGO and
 VIRGO. The detection of these gravitational waveforms (GW) is important for
 testing general relativity in the strong field limit. Moreover, these 
detectors can extract from the waves physical data about these sources such as 
the component masses and spins and the orbital eccentricity. For an unbiased
 extraction of these parameters, a large bank of accurate waveforms needs to be
constructed. Numerical relativity alone cannot compute all the waveforms needed
 because of the computational cost. Instead, the waveforms are based on 
post-Newtonian (PN) approximations~\cite{Blanchet04,Blanchet2006}.
 
The post-Newtonian approximation is a slow-motion, weak-field approximation to
general relativity. In order to produce a post-Newtonian waveform, the PN 
equations of motion of the binary are solved to yield explicit expressions for
the accelerations of each body in terms of the binary's orbital frequency 
$\Omega$~\cite{Jaranowski98a, Jaranowski99a, Damour00a, Damour01a, 
Blanchet00a, Blanchet01a, Damour01b, Blanchet04, Itoh01, Itoh03, Itoh04}.
Then solving the post-Newtonian wave generation problem yields expressions
for the gravitational waveform $h$ and the gravitational wave flux $F$ in terms
of radiative multipole moments~\cite{thorne80}. These radiative multipole 
moments are in turn related to the source multipole moments, which can be given
in terms of the relative position and relative velocity of the 
binary~\cite{Blanchet98}. Instead of comparing the post-Newtonian waveform with
 a numerical waveform along a certain direction with respect to the source, 
the comparison can be done in all directions by decomposing the waveform in
terms of spherical harmonic modes. For an equal-mass non-spinning binary, the 
$(2,2)$ mode $h_{22}$~\cite{Kidder07a,ABIQ,ABIQ-erratum,BFIS} is often used to compare numerical and 
post-Newtonian waveforms because it is the dominant mode. Its time derivative 
$\dot{h}_{22}$ is used to compute the gravitational wave flux. The resulting 
expressions for the orbital energy $E$, the gravitational energy flux $F$ and
 the amplitude $h_{22}$ are given as Taylor series of the frequency-related 
parameter 
\begin{equation}
x=(M \Omega)^{2/3}\, ,
\end{equation}
where  $M$ is the total mass of the binary and $G=c=1$. The invariantly defined
``velocity''
\begin{equation}
 v=x^{1/2}\, ,
\end{equation}
another dimensionless parameter, is often used in writing these Taylor series.

\begin{figure}
%\psfrag{n}{\large $n$}%\footnotesize \scriptsize \tiny
%\includegraphics[scale=0.47]{convergence-test-exp}
\includegraphics[scale=0.47]{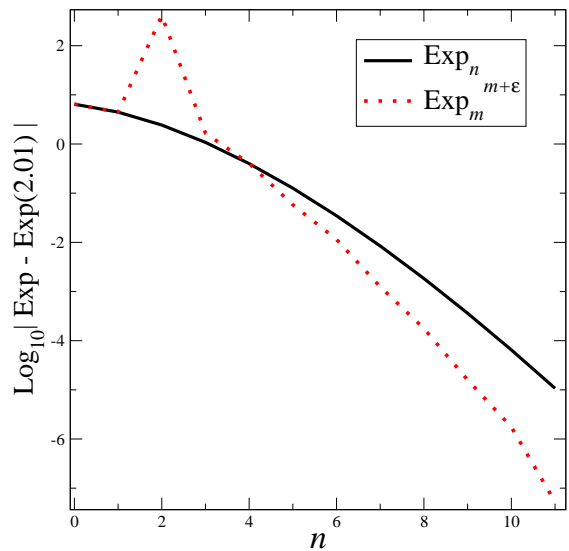}
\caption{\label{fig:exp-convergence} Convergence of the Taylor expansion, 
${\rm Exp}_n=\sum_{k=0}^{n} v^k/k!$ of the exponential function ${\rm Exp}(v)$
 and its Pad\'e approximant ${\rm Exp}_{m}^{m+\epsilon}$ at $v=2.01$, 
$m=\lfloor n/2\rfloor$. The Pad\'e approximant converges 
faster than the Taylor series.}
\end{figure}

Computing PN series to high order is difficult and time consuming. Since the
 various PN expressions are given as slowly convergent Taylor series, the 
Pad\'e transformation~\cite{BenderOrszag,nr3} was suggested in 
Ref.~\cite{Damour98} to accelerate the convergence of these series. The Pad\'e 
transformation, $P^m_n$, consists of writing a Taylor series, $T_k$, of order
 $k$ as the ratio of two polynomials, one of order $m$ in the numerator, and 
another of order $n$ in the denominator, such that $m+n \le k$. If 
well-behaved, this method accelerates the convergence of a Taylor series as the
 order of the Pad\'e transformation, $m+n$, is increased. For example, in 
Table~\ref{tab:taylor-pade-exponential} we compare the convergence of the 
Taylor expansion of the exponential function, 
${\rm Exp}_n(v)(\equiv \mathbf{\mathit{e}}^v)$,  at order $n$ to its Pad\'e 
approximant, ${\rm Exp}_m^{m+\epsilon}(v)=P_m^{m+\epsilon}[{\rm Exp}_n(v)]$,
 along the diagonal, where $m={\rm \lfloor}n/2\rfloor$ and $\epsilon=0$ or $1$.
After twelve terms $(n=11)$, the last two partial sums of the Taylor expansion
 converge to $4$ significant figures. However, the last two Pad\'e approximants
 ${\rm Exp}_5^5(v)$ and ${\rm Exp}_5^6(v)$ converge to $6$ significant figures.
 The error between the exact value of the exponential, $7.46331734$, and the
 Pad\'e approximant ${\rm Exp}_5^6(v=2.01)$ is $6\times 10^{-8}$, while the
 error between the eleventh order partial sum and the exact value is $10^{-5}$.
Fig.~\ref{fig:exp-convergence} shows the convergence of the Taylor expansion 
of the exponential function and its Pad\'e approximant. 
 
\begin{table}
\begin{tabular}{c|c|c}
n & ${\rm Exp}_n(v)$ & $P_m^{m+\epsilon}[{\rm Exp}_n(v)] $ 
\\\hline
0&	1.0000000&	1.0000000
\\1&	3.0099999&	3.0099999
\\2&	5.0300499&	-401.0000
\\3&	6.3834834&	9.1313636
\\4&	{\bf 7}.0635838&	{\bf 7}.0601492
\\5&	{\bf 7}.3369841&	{\bf 7.4}053299
\\6&	{\bf 7.4}285732&        {\bf 7.4}747817
\\7&	{\bf 7.4}548724&	{\bf 7.46}45660
\\8&	{\bf 7.46}14801&	{\bf 7.463}1404
\\9&	{\bf 7.46}29558&	{\bf 7.4633}014
\\10&	{\bf 7.463}2524&	{\bf 7.46331}91
\\11&	{\bf 7.463}3066&	{\bf 7.46331}74
\end{tabular}
\caption{\label{tab:taylor-pade-exponential} 
Convergence of the Taylor expansion, 
${\rm Exp}_n=\sum_{k=0}^{n} v^k/k!$ of the exponential function ${\rm Exp}(v)$
 and its Pad\'e approximant ${\rm Exp}_{m}^{m+\epsilon}$ at $v=2.01$, 
$m=\lfloor n/2\rfloor$. The Pad\'e approximant converges to six significant 
figures while the Taylor series converges to four significant figures at
 $v=2.01$. The error between the exact value of the exponential, $7.46331734$,
 and the Pad\'e approximant ${\rm Exp}_5^6(v=2.01)$ is $6\times 10^{-8}$, while
 the error between the Taylor approximant ${\rm Exp}_{11}(v=2.01)$ and the
 exact value is $10^{-5}$.}
\end{table}

%%%%%%%%%%%%%%%%%%%%%%%%%%%%%%%%%%%%%%%%%%%%%%%%%%%%%%%%%%%%%%%%
%%%%%%%%%%%%%%%%%%%%%%%%%%%%%%%%%%%%%%%%%%%%%%%%%%%%%%%%%%%%%%%%

The hope of accelerating the convergence of the post-Newtonian Taylor series of
 the energy and flux motivated the use of their Pad\'e approximants to 
construct Pad\'e approximant waveforms~\cite{Damour98,Damour2001,Damour02,2000PhRvD..62h4011D,Damour2007,Damour2007a,Buonanno:2002ft,Buonanno00,Buonanno06,Buonanno2007,Buonanno99,DamourJena2008,Porter2005}. 
 If these resummation techniques accelerate the convergence of the Taylor 
series in PN approximations, the range of validity of PN 
approximations suggested by Ref~\cite{berti-yunes} could be extended. 
Moreover, the work of Refs.~\cite{Poisson:1993vp,Cutler93} in the test mass 
limit motivated the addition of a simple pole to the flux $F$ of a binary 
system as the bodies approach the light ring orbit. By mathematical continuity,
the existence of a pole in the equal mass case was anticipated~\cite{Damour98}.

More recently, waveforms are constructed by including these ideas in Effective
 One Body (EOB) models. The EOB approach~\cite{Buonanno00,Buonanno06,Buonanno2007,Buonanno99,Buonanno:2002ft,Buonanno1994,Damour01c,Damour02,Damour03,Damour2001,Damour2008,Damour:2007cb,DamourJS2008,Damour2007,Damour2007a,Damour06,DN2007b}
aims at providing an accurate analytical description of the motion and 
radiation of coalescing binary black holes. The approach consists of three 
separate ingredients: 1) a description of the conservative Hamiltonian
part of the dynamics $\hat{H}$, 2) a formulation of the radiation reaction 
force $\cal F$ from the radiated flux $F$ and 3) an expression of the GW 
waveform amplitude emitted by the coalescing binary system (i.e $h_{22}$).

The flux plays an important role in approximating the radiation reaction force
 $\cal F$ in the EOB models~\cite{Iyer93,Iyer95,Buonanno00}. The leading-order 
radiation reaction force $\cal F$~\cite{Damour1981,Damour83book,Schafer86book}
 enters the equations of motion at 2.5PN order. Since the equations of motion
 are known only to 3.5PN order, one has to rely on the assumed balance between
 energy loss in the system and radiated flux at 
infinity~\cite{Blanchet95b,Blanchet97} to generate an approximate expression of
the radiation reaction force at 3.5PN order beyond the leading term. 

 Ref.~\cite{Toprep} computes the GW energy flux and GW
frequency derivative from a highly accurate numerical simulation of an 
equal-mass, non-spinning black hole binary. By assuming energy balance, the 
(derivative of the) center-of-mass energy is estimated. These quantities 
are then compared to the numerical values using various Taylor, Pad\'e and EOB
models.  The main goal of 
Ref.~\cite{Toprep} is taking a set of well-established proposals in the 
literature for approximating waveforms and seeing how well they work in 
practice. Another goal of Ref.~\cite{Toprep} is to examine some modifications 
of those proposals. The main goal of this paper, by contrast, is to show that 
a key ingredient in those proposals does not appear to be necessary.

In Ref.~\cite{Blacnchet2002-ISCO}, Blanchet gave an argument that Pad\'e and EOB
resummations are unjustified because for two comparable-mass bodies there
is no equivalent of the Schwarzschild light-ring orbit at the radius $r=3M$.
His argument is based on the PN coefficients of the binary's
energy and their relation to predicting the innermost circular orbit. He finds
that the radius of convergence of the PN series, which is
related to the radius of the light-ring orbit, is  around
one (instead of $1/3$ as for
Schwarzschild). Blanchet concluded that Taylor series converge well for
equal masses and that templates based on Pad\'e/EOB
are not justified because the dynamics of two bodies in General Relativity
does not
appear as a small "deformation" of the motion of a test particle in
Schwarzschild. This paper arrives at similar conclusions but not by considering
the innermost circular orbit, which is not precisely defined in the full
nonlinear case. Instead, we compare
Pade approximants of the flux and Pade/EOB waveforms to the numerical data
of Refs.~\cite{Poisson95,Boyle2007}.

In this paper, we focus on testing two main techniques involved in building EOB
 models: the systematic use of Pad\'e approximants, and the addition of a pole
 to the flux. The  goal is to simplify these models by removing any unnecessary
 procedures in designing waveforms that provide good agreement with numerical 
waveforms. 

  Damour et al.~\cite{Damour98,Damour2001} first suggested  techniques for 
resumming the Taylor expansions of the energy and flux functions. Starting from
 the PN expansions of the energy $E$ and the flux $F$, they proposed a new 
class of waveforms called $P$-approximants, based on three essential 
ingredients. The first step is the introduction of new energy-type
 (Eq.~\ref{eq:e}) and flux type (Eq.~\ref{eq:NewFlux}) functions,
 called $e(v)$ and $f(v)$ respectively. The second step is to 
Pad\'e-approximate the Taylor expansion of these functions. The third step is
 to use these Pad\'e transforms in the definition of the energy $E$
 (Eq.~\ref{eq:Energy}) and Pad\'e-approximated flux (Eq.~\ref{eq:fourthFlux}). 
The last step is to construct either the Pad\'e-approximated waveform as in 
Sec.~\ref{sec:waveforms} or the EOB waveform as in Sec.~\ref{sec:simpleeob}. 
Schematically, the suggested procedure is summarized by the following map:

%\begin{widetext}
\begin{eqnarray}
 \bigg[E_{n} , F_{n}\bigg] \rightarrow  \bigg[e_{n} , f_{n}\bigg] \rightarrow  \bigg[e_n^m ,
f_n^m\bigg]
 \nonumber \\ 
\rightarrow \bigg[E(e_n^m) , F(f_n^m)\bigg] 
\rightarrow h \, . 
\label {eq:map}
\end{eqnarray}
%\end{widetext}

Our notation is to denote by $T^m_n(x)$ the Pad\'e approximant of 
a $k$-th order Taylor series $T_k(x)$ with an $m$-th order polynomial  
in the numerator and an $n$-th order polynomial in the denominator 
such that $m+n\leq k$, i.e.
the Pad\'e approximant of $e_{k}(x)$ is $e^m_n(x)$.

 In Section~\ref{sec:energy}, we compare the 3PN Taylor series of the energy
 function to its possible Pad\'e approximants using the intermediate energy
 function $e(x)$, as suggested by Damour et al.~\cite{Damour98}. We compute the
 last stable orbit frequency, defined as the frequency for which the energy 
reaches a minimum as a function of frequency, and also the poles of the energy in
 the complex plane corresponding to each possible Pad\'e approximant. The large
 variation of last stable orbit frequency and poles does not suggest good
 convergence of the Pad\'e-approximated intermediate energy function $e(x)$.
 The energy function $E(x)$ is strongly dependent on the choice of the Pad\'e 
approximant of $e(x)$. Accordingly, the Pad\'e waveform will also be strongly
 dependent on the choice of the Pad\'e approximant.  

In Section~\ref{sec:flux}, we present two possible methods for calculating the
 Pad\'e approximant of the flux function. The first method simply takes the 
Pad\'e approximant of the Taylor series 
treating the logarithmic contribution as constant. 
Following~\cite{Damour98}, the second method adds a pole
to the Taylor series, factors out the logarithmic contribution to the series, 
and then computes the Pad\'e approximant of the resulting Taylor series. 
 We test the convergence
of the Pad\'e approximant for both methods versus their Taylor series. 
We find that the Pad\'e approximants of the flux do not converge any faster 
 than their Taylor counterpart.

A simple example that illustrates the problem is shown in 
Table~\ref{tab:summaryFactorizedFlux}. There we compare the partial sums
of the Taylor series for the flux with the corresponding Pad\'e approximants 
in the test mass limit. The four flux functions $\overline{F}_n$, 
$\overline{F}_m^{m+\epsilon}$, $F_n$ and $F_m^{m+\epsilon}$ are given in 
Eqs.~\ref{eq:firstFlux}, \ref{eq:secondFlux}, \ref{eq:thirdFlux} and 
\ref{eq:fourthFlux} respectively. Even for a relatively small value of $x$, 
namely
 $x=0.04$ $(v=0.2)$, the Taylor series is converging very slowly. After 12 
terms, only about 4 or 5 significant digits seem reliable. Moreover, the Pad\'e
 resummation shows very similar behavior; there is no improvement in the
 convergence. We will return to this example in 
Fig.~\ref{fig:TestMass-FluxConvergence}.

\begin{table}
\begin{tabular}{c|c|c|c|c}
PN order &$\overline{F}_n$  & $\overline{F}_m^{m+\epsilon}$  &$F_n$ 
&$F_m^{m+\epsilon}$
\\\hline
0.0&	1.000000&	1.000000&	1.530011&	1.530011
\\
0.5&	1.000000&	1.000000&	1.000000&	1.000000
\\
1.0&	0.851547&	1.000000&	0.772866&	0.602534
\\
1.5&	{\bf 0.9}52078&	{\bf 0.9}11487&	1.005361&	0.887757
\\
2.0&	{\bf 0.9}44193&	{\bf 0.9}28720&	{\bf 0.9}40013&	{\bf 0.93}7227
\\
2.5&	{\bf 0.9}31939&	{\bf 0.93}6461&	{\bf 0.9}25444&	{\bf 0.93}8929
\\
3.0&	{\bf 0.9}41025&	{\bf 0.939}366&	{\bf 0.9}45405&	{\bf 0.93}9502
\\
3.5&	{\bf 0.939}726&	{\bf 0.939}399&	{\bf 0.93}8991&	{\bf 0.93}8082
\\
4.0&	{\bf 0.939}208&	{\bf 0.939}363&	{\bf 0.939}048&	{\bf 0.939}471
\\
4.5&	{\bf 0.939}745&	{\bf 0.939}719&	{\bf 0.939}979&	{\bf 0.939}516
\\
5.0&	{\bf 0.93960}1&	{\bf 0.9396}53&	{\bf 0.939}526&	{\bf 0.9396}84
\\
5.5&	{\bf 0.93960}5&	{\bf 0.9396}23&	{\bf 0.939}616&	{\bf 0.9396}21
\end{tabular}
\caption{\label{tab:summaryFactorizedFlux} Convergence of the Taylor series and
its Pad\'e aprroximants of the flux in the test particle limit at $v=0.2$ 
($x=0.04$). The four flux functions $\overline{F}_n$, 
$\overline{F}_m^{m+\epsilon}$, $F_n$ and $F_m^{m+\epsilon}$ are given in 
Eqs.~\ref{eq:firstFlux}, \ref{eq:secondFlux}, \ref{eq:thirdFlux} and 
\ref{eq:fourthFlux} respectively. Even in the test mass limit, the Pad\'e 
approximant of the flux 
fails to converge faster that its $5.5$ PN Taylor series at a relatively small
 value of $v=0.2$.  After $12$ terms, only about 4 or 5 significant digits 
seem reliable for the Taylor expansions and their Pad\'e approximants. The 
lack of improvement in the convergence of the Pad\'e approximants should be 
contrasted with the example in  
Table~\ref{tab:taylor-pade-exponential}.}
\end{table}

In Section~\ref{sec:waveforms}, we generate all the possible Pad\'e 
waveforms as 
suggested by Damour et al.~\cite{Damour98} corresponding to 3 and 3.5 PN order.
 The waveform approximation requires the choice of a pole. We use 
the only physical pole, found from the 2PN Pad\'e-approximated energy 
$E^1_1$. We also use the last stable orbit from   
the 3PN energy Taylor series $E_3$. The results are not very sensitive to 
this choice. We compare the Pad\'e waveforms 
to a 15-orbit numerical waveform in the equal mass, nonspinning quasi-circular 
case~\cite{Boyle2007}. 
 The phase difference in these comparisons ranges between 0.05 
and a few radians for well-defined Pad\'e approximants (not having a pole in 
the frequency domain of interest) when the matching of the numerical and 
Pad\'e waveforms is done at the gravitational 
wave frequency 
$M\omega=0.1$~\cite{Boyle2007}. None of the Pad\'e waveforms agrees with the 
numerical waveform better than the Taylor series T4-3.5/3.0PN, which has an 
error of 0.02 radians. (We identify post-Newtonian approximants with three
  pieces of information: the label introduced by~\cite{Damour2001} for
  how the orbital phase is evolved; the PN order to which the orbital
  phase is computed; and the PN order at which the amplitude of the
  waveform is computed.  See Ref.~\cite{Boyle2007} for
  more details.) Our conclusion
is that the Pad\'e approximant might be helpful in suggesting fitting formulas
but it does not provide a more rapidly convergent method. Note that the Pad\'e 
transform also fails to accelerate the convergence of the T2, T3 and $h_{22}$ 
Taylor series (see Refs.~\cite{Boyle2007,Damour2001} for the definition of these 
Taylor series).

In Section~\ref{sec:simpleeob}, based on the results of the previous sections,
 we design a simple EOB model (closely related to the ET EOB model of 
Ref.~\cite{Buonanno:2002ft}) using the Taylor series of the flux. We add one
 unknown 4PN term that we fit for by maximizing the agreement between the EOB
 model waveform and the numerical waveform. The model does not require adding 
a pole to the flux, nor an a priori knowledge of the last stable orbit from the
 energy function. This simple EOB model, with only one parameter to fit for,
 agrees with the numerical waveform to within 0.002 radians ($3\times 10^{-4}$ 
cycles). (This is six times
 smaller than the claimed numerical accuracy of~\cite{Damour2007a}, 
smaller by an even larger factor than the claimed numerical accuracy 
of~\cite{DamourJena2008}, and 
 25 times smaller than the gravitational wave phase uncertainty of the
 numerical waveform. See Table III in Ref~\cite{Boyle2007} for more details.)
 This model agrees with the numerical waveform better than any previously
 suggested Taylor, Pad\'e or EOB waveform.

%###############################################
\section{Energy Function}
%###############################################
\label{sec:energy}

Damour et al.~\cite{Damour98} introduced a new energy-type function 
$e(x)$, where $x$
is the PN frequency related parameter. This assumed more ``basic'' energy
 function $e(x)$
 is constructed out of the total relativistic energy $E_{\rm tot}(x)$ of the 
binary system. Explicitly
\begin{equation}
\label{eq:e}
e(x)  \equiv \left( \frac{E_{\rm tot}^2 - m_1^2
-m_2^2}{2m_1 m_2}\right)^2 -1 \,, 
\end{equation}
where $m_1$, $m_2$ are the masses of the bodies.
The total relativistic energy function $E_{\rm tot}$ is related to the post 
Newtonian energy function $E(x)$ through
\begin{equation} 
\label{eq:Etot}
E_{\rm tot}(x)=M\left[1+E(x)\right]\,,
\end{equation}
where $M$ is the total mass ($M=m_1+m_2$).
Solving for $E(x)$ in terms of $e(x)$ using Eqs.~(\ref{eq:e}) and 
(\ref{eq:Etot}), we get\cite{Damour98}
\begin{equation}
\label{eq:Energy}
E(x)=\left\{1+2 \nu \left[\sqrt{1+e(x)}-1\right]\right\}^{1/2}-1 \,,
\end{equation}
where the symmetric mass ratio is $\nu=m_1m_2/M^2$.
The orbital energy function $E(x)$ is known as a Taylor series $E_k$ up to 3PN
order as a function of $x$ and $\nu$~\cite{Blanchet2006}
%\begin{widetext}
\begin{align}
\label{eq:E3PN}
E_{\rm 3PN} (x) =& -\frac{1}{2} \, \nu \, x \, \bigg\{ 1-\frac{1}{12} \,
(9+\nu) \, x \nonumber \\ &
\qquad \qquad -\frac{1}{8} \, \big( 27 - 19\nu  
+ \frac{1}{3}\nu^2 \big) \, x^2  
 \nonumber \\ & \qquad \qquad
+\Big[-\frac{675}{64}+ \left(\frac{34445}{576}-\frac{205}{96} \pi^2\right) \nu 
\nonumber \\ &  \qquad \qquad \quad \,\,\,
-\frac{155}{96}\nu^2 
-\frac{35}{5184}\nu^3 \Big] \, x^3
\bigg\} \, . 
\end{align}
%\end{widetext}
Using the above equations, we compute the Taylor series 
expansion, $e_k(x)$, of $e(x)$ up to 3PN order:
%\begin{widetext}
\begin{align}
e_{\rm 3PN}(x) 
=&
-x \bigg\{1-(1+\frac{1}{3}\nu)x-(3-\frac{35}{12}\nu)x^2
\nonumber \\
& \qquad 
- \Big[ 9 +\frac{1}{288} \left(-17236+615 \pi ^2\right) \nu 
\nonumber \\ 
& \qquad \quad \,\,\,
+\frac{103}{36}\nu ^2 -\frac{1}{81}\nu ^3\Big]\, x^3
\bigg\} \, .
\label{eq:eTpn}
\end{align}
%\end{widetext}
In the test mass limit ($\nu\rightarrow0$), the exact function $e(x)$ coincides with the 
Pad\'e approximant $P^1_1(x)$ of its Taylor expansion in Eq.~(\ref{eq:eTpn})
\begin{equation}
e(x;\nu\rightarrow0)=-x\frac{1-4x}{1-3x} \,.
\end{equation}
This quantity has a pole at $x_{\rm pole}=1/3$. The orbital energy is then  
\begin{equation}
E(x;\nu \rightarrow0)= \nu \left(\sqrt{1-x\frac{1-4x}{1-3x}}-1\right) \,,
\end{equation}
and it derivative is
\begin{equation}
\frac{dE(x;\nu \rightarrow0)}{dx}=-\nu \frac{1-6x}{2(1-3x)^{3/2}} \,.
\end{equation}
The last stable orbit occurs where 
\begin{equation}
\label{eq:lso}
\frac{dE}{dx}=0  \,, 
\end{equation}
so in the limit $\nu\rightarrow0$ the last stable orbit is at exactly 
$x_{\rm lso}=1/6$. On the grounds of mathematical continuity 
between the test mass limit $\nu\rightarrow0$ 
and the finite mass ratio case, Damour et al.~\cite{Damour98} argued that the
 exact function $e(x)$ should be meromorphically extendable in at least part of
 the complex plane and should have a simple pole on the real axis. They 
suggested that Pad\'e approximants would be excellent tools for giving accurate
representations of functions having such poles.

%%%%%%%%%%%%%%%%%%%%%%%%%%%%%%%%%%%%%%%%%%%%%%%%%%%%%%%%%%%%%%%%
\begin{figure}
\psfrag{x}{\large $x$}%\footnotesize \scriptsize \tiny
\includegraphics[scale=0.47]{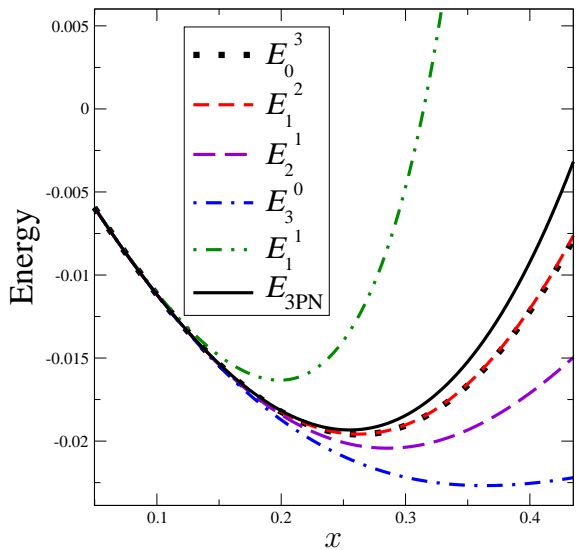}
\caption{\label{fig:Energy} Post Newtonian Energy at 3PN and its Pad\'e 
approximants for the case $\nu=1/4$. The plot includes the high value 
of $x_{\rm lso}=0.36$, the numerical 
data available is at $x=0.16$. The plots of $E^3_0$, $E^2_1$,
  $E^1_2$ and $E^0_3$ vary significantly although they all correspond 
to the 3PN Taylor series of the energy function. $E^1_1$ is very 
different from the other functions, which suggests a poorly convergent Pad\'e 
approximant.}
\end{figure}
%%%%%%%%%%%%%%%%%%%%%%%%%%%%%%%%%%%%%%%%%%%%%%%%%%%%%%%%%%%%%%%%

Once we know the Taylor series of the new energy function, $e_k(x)$, we compute
 its Pad\'e approximant $e_n^m(x)$, with $m+n\leq k$. The Pad\'e-approximated 
energy, $E_n^m(x)$ is obtained by replacing $e(x)$ in Eq.~(\ref{eq:Energy}) 
with $e_n^m(x)$. In the equal mass case ($\nu=1/4$), we can define several 
Pad\'e approximants of $e_k(x)$. The most interesting Pad\'e approximants have 
a maximal sum of their indices, since they should be closest to 
the unknown exact function if the Pad\'e resummation is converging. 
In Fig.~\ref{fig:Energy}, we show a plot of the PN energy function 
$E_{\rm 3PN}(x)$ and its Pad\'e approximants $E_1^1$, 
$E_1^2$, $E_2^1$, $E_3^0$ and $E_0^3$ as a function of $x$.

Although the Pad\'e approximants of the energy are of maximal order, 
they differ significantly. Good convergence of the Pad\'e approximants 
requires 
 good agreement between approximants of the same order $n+m$, if there is 
no pole in the region of interest ($0<x\lesssim 0.4$).
For example, there is no a priori
reason why one should prefer either $E_1^2$ or $E_2^1$. Although both 
have the same order and are equally close to the diagonal, the difference  
 between these functions is quite large. 

In Table~\ref{tab:polelso} we compute the locations of the poles and the last 
stable orbits for all of these Pad\'e approximants.
The ill-convergence of the Pad\'e transform is again seen by looking  
at the variation of the last stable orbit positions. In Table~\ref{tab:polelso}
 for example,
$x_{\rm lso}$ of $E_1^2$ differs by about $8\%$ from $x_{\rm lso}$ of $E_2^1$.
Moreover, for finite $\nu$, the poles are all complex or not in
 the interval $[0,1]$ except for the case $x_{\rm pole}=52/109$, corresponding
 to the Pad\'e-approximated energy $E_1^1$. There is no reason why this should
 be the ``exact'' pole that should be used in the formalism, since none of the 
third-order Pad\'e approximants of the 3PN energy has a physical pole.

In summary, using Pad\'e approximants for the 
energy function in the equal mass case does not seem to provide any benefit.
The differences between the various Pad\'e approximants of the energy are 
large. The quantities $x_{\rm pole}$ and $x_{\rm lso}$ do not show any regular
 behavior that could be a sign of a  physical pole that could be found by using
 the Pad\'e transform.

%\hspace{1cm}

%###############################################
\section{Flux Function}
%###############################################
\label{sec:flux}

The general form of the PN flux at order $N$ is:
\begin{eqnarray}
\label{eq:Flux}
F(v) =\frac{32}{5} \nu^2 v^{10} \times \overline{F}_N \, ,
\end{eqnarray}
where the normalized flux $\overline{F}$ is a Taylor expasion in $v$ with 
logarithmic terms
\begin{equation}
\label{eq:firstFlux}
\overline{F}_N(v) =\sum_{k=0}^{2N}A_k v^k +
\left(\sum_{k=6}^{2N} B_k v^k\right) \log v  \, ,
\end{equation} 
where the post-Newtonian coefficients $A_i$ and $B_i$ are functions of the mass
ratio parameter $\nu$. They are given in the test mass limit in 
Ref.~\cite{Tanaka96} and in the equal mass quasi-circular case
in Ref.~\cite{Blanchet2006}.
 The flux series has a logarithmic contribution starting at 3PN. 
Pad\'e approximants, however, are well defined only for pure polynomials.
 Two possible methods are therefore used to compute the Pad\'e approximant of
 the flux. The first method simply treats the logarithmic terms as constants
 and resums the series as a pure polynomial such that the Pad\'e-approximated
 flux $\overline{F}_n^m$ is 
\begin{equation}
\label{eq:secondFlux}
\overline{F}_n^m(v) = P_n^m \left[\, \overline{F}_N(v) \, \right]  \,.
\end{equation}

The second method, suggested by Ref.~\cite{Damour98}, defines a new flux 
function $f$ by adding a pole, factoring the logarithmic terms from the series,
 and finally computing the  Pad\'e approximant of the pure polynomial. Since
 we would like to check the convergence of the Pad\'e-approximated flux versus
 its Taylor series, we sketch the definitions of the various 
functions involved. According to Ref.~\cite{Damour98}, two ideas are needed for
a good representation of the analytic structure of the flux. First, since in
 the test mass limit $F$ is thought to have a simple pole at the light ring~\cite{Cutler93},
 one might  expect it by continuity to have a pole in the comparable mass case.
 This motivates the introduction of the following factored flux function,
 $f (v;\nu)$:
\begin{equation}
\label{eq:NewFlux}
f(v;\nu)\equiv \left( 1-\frac{v}{v_{\rm pole}(\nu)} \right) F(v;\nu)\, ,
\end{equation}
where $v_{\rm pole}$ is the pole of the Pad\'e-approximanted energy function 
used.

\begin{table}
\begin{tabular}{c|c|c}
Energy       &  $x_{\rm pole}$                 & $x_{\rm lso}$
\\
\hline
$E_{\rm 3PN}$&  $-$                            & $0.254$
\\
$E_1^1$      & $52/109=0.477$                  & $0.199$ 
\\
$E^3_0$      &  $-$                            & $0.262$ 
\\  
$E^2_1$      &  $-4.41$                        & $0.261$ 
\\
$E^1_2$      &  $0.170 \pm  0.757i  $          & $0.285$
\\ 
$E^0_3$      &  $0.044 \pm  0.501i$,$-0.696$   & $0.363$
\end{tabular}
\caption{\label{tab:polelso} Values of the poles and last stable 
orbit (lso) of the energy for the case $\nu=1/4 $.
 The poles $x_{\rm pole}$ and last stable 
orbit frequency of the function 
$E_n^m(x)$ depend significantly on which Pad\'e approximant is constructed 
from the Taylor series  $e_k(x)$. The only physical pole is 
 $x_{\rm pole}=52/109$, which is at a larger value than the pole 
 in the test mass limit. The position of the last stable orbit also varies 
significantly.}
\end{table}

Second, the logarithmic term that appears in the flux
function needs to be factored out so we can use the standard Pad\'e 
transformation. After factoring the logarithmic terms out, the flux function 
$f$ becomes
%\begin{widetext}
\begin{eqnarray}
\label{eq:FactorizedNewFlux}
 f_{n} (v; \nu ) &=&  \left[ 1+ \log\frac{v}{v_{\rm lso}} \left( \sum_{k=6}^{2N} \ell_k v^k \right)\right] 
 \nonumber \\ 
&&
\times 
\left(\sum_{k=0}^{2N} f_k v^k\right) \, ,
\end{eqnarray}
%\end{widetext}
where the coefficients $l_k$ and $f_k$ are given in Ref.~\cite{Damour98}, and
 $v_{\rm lso}$ is the velocity of the last stable orbit of the 
Pad\'e-approximated energy. Then the Taylor series of the flux with a pole is 
defined as 
\begin{equation}
\label{eq:thirdFlux}
F_n(v;\nu)\equiv \frac{f_n(v;\nu)}{1-v/v_{\rm pole}(\nu)}
\, . 
\end{equation}
The Pad\'e approximant of the intermediate flux function
 $f(v)$ is defined as
%\begin{widetext}
\begin{eqnarray}
f_n^m (v) &\equiv&  
\left[ 1+\log\frac{
v}{v_{\rm lso}(e_n^m;\nu)} \, \left( \sum_{k=6}^{2N} \ell_k v^k \right) \right] 
\nonumber \\ && 
\, \times P_n^m \left[ \sum_{k=0}^{2N} f_k \, v^k \right] \, ,
\label{eq:intermediatefluxpade}
\end{eqnarray}
%\end{widetext}
where $v_{\rm lso}(e_n^m;\nu)$ denotes the last stable orbit velocity  for the
  Pad\'e approximant $P_n^m\big[e(x)\big]$. Finally, the corresponding Pad\'e
 approximant of the flux $F(v)$ is given by 
\begin{equation}
\label{eq:fourthFlux}
F_n^m(v;\nu)\equiv \frac{f_n^m(v;\nu)}{1-v/v_{\rm pole}(e_n^m;\nu)}
\, , 
\end{equation}
where $v_{\rm pole}(e_n^m;\nu)$ denotes the pole velocity defined by 
$e_n^m(x)$.

%%%%%%%%%%%%%%%%%%%%%%%%%%%%%%%%%%%%%%%%%%%%%%%%%
\subsection{Flux for the test mass case }
%%%%%%%%%%%%%%%%%%%%%%%%%%%%%%%%%%%%%%%%%%%%%%%%%

The exact gravitational wave luminosity $F$ is not known analytically in the
 test particle limit. It has been computed numerically by 
Poisson~\cite{Poisson95}. The post-Newtonian expansion of the flux is known in 
the test mass limit to 5.5PN order~\cite{Tanaka96}. This allows us to test the
 rate of convergence of the Taylor series of the normalized flux 
$\overline{F}_n$ (Eq.~\ref{eq:firstFlux}) and its Pad\'e-approximant  
$\overline{F}_n^m$ constructed treating the logarithmic term as a constant 
(Eq.~\ref{eq:firstFlux}). We also test the convergence of the
flux function $F_n$ (Eq.~\ref{eq:thirdFlux}) and its Pad\'e 
approximant $F_n^m$ (Eq.~\ref{eq:fourthFlux}).
These convergence tests use the known values $v_{\rm pole}=1/\sqrt{3}$ and 
$v_{\rm lso}=1/\sqrt{6}$ for the test mass limit as discussed in 
Sec.~\ref{sec:energy}.

%%%Converegence test along the sup-diagonal terms

\begin{figure}
\includegraphics[scale=0.47]{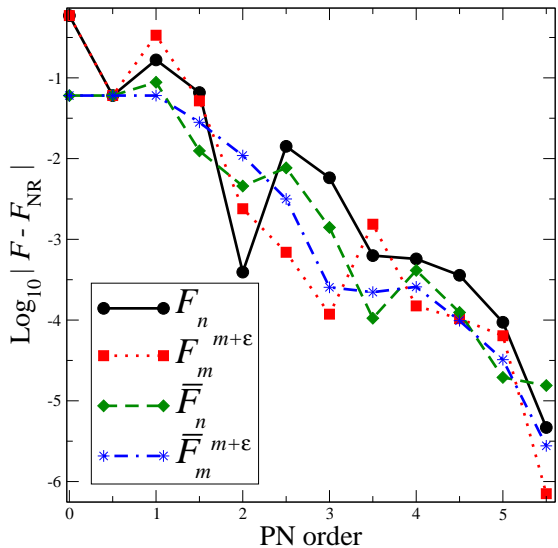}
\caption{\label{fig:TestMass-FluxConvergence} 
Convergence of the flux approximations in the test
 mass limit for $v=0.2$. The four flux functions $\overline{F}_n$, 
$\overline{F}_m^{m+\epsilon}$, $F_n$ and $F_m^{m+\epsilon}$ are given in 
Eqs.~\ref{eq:firstFlux}, \ref{eq:secondFlux}, \ref{eq:thirdFlux} and 
\ref{eq:fourthFlux} respectively. The Pad\'e approximants do not converge 
faster than their Taylor series counterparts. The Pad\'e and Taylor series 
alternate at providing the best agreement with the  ``exact'' data as the PN
 order increases. Contrast the behavior here with 
Fig.~\ref{fig:exp-convergence}.}
\end{figure}

In Fig.~\ref{fig:TestMass-FluxConvergence}, we test the convergence of the 
various flux functions at the velocity value $v=0.2$. The four flux functions 
$\overline{F}_n$, $\overline{F}_m^{m+\epsilon}$, $F_n$ and $F_m^{m+\epsilon}$
 are given in Eqs.~\ref{eq:firstFlux}, \ref{eq:secondFlux}, \ref{eq:thirdFlux} and 
\ref{eq:fourthFlux} respectively.
We use the Pad\'e approximant 
along the diagonal $P_m^{m+\epsilon}$ where $\epsilon=0$ or $1$.
The rates of 
convergence of the Taylor expansion and its Pad\'e approximant are nearly 
equal for the two methods, whether or not we include a pole.
 As the PN
 order increases, the Taylor series and its Pad\'e approximant alternate in which
 provides a better fit
 to the numerical data for the flux. For example, at 2PN order the Taylor 
flux
with a pole (Eq.~\ref{eq:thirdFlux}) fits the numerical data the best.
At 2.5 and 3 PN order the Pad\'e approximant of the flux $F_n^m$ 
(Eq.~\ref{eq:fourthFlux})
 fits the numerical data the best, while at 3.5 and 5PN order the Taylor series
 of the flux (Eq.~\ref{eq:firstFlux}) is the best. At 5.5PN the Pad\'e 
approximant of the flux 
(Eq.~\ref{eq:fourthFlux}) gives the best agreement.
The results are similar for other values of $v$. No method has the best
 convergence rate.

%%%Comparing the maximal Pade terms

\begin{figure}
\psfrag{x}{\large$n$}
\psfrag{v }{$v$}
\includegraphics[scale=0.47]{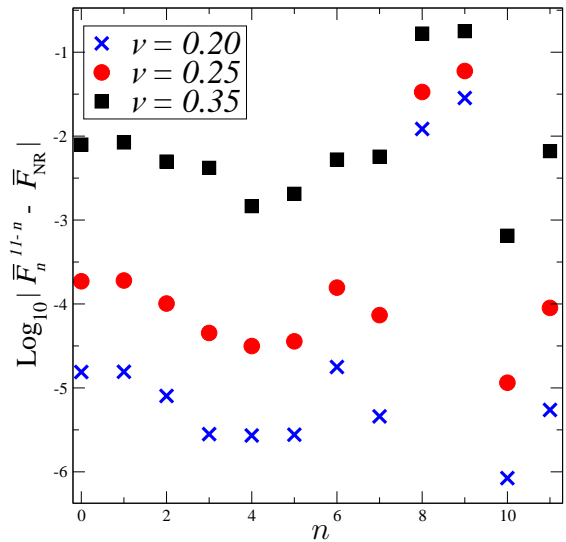}
\caption{\label{fig:TestMass-maximalpade}Error between maximal Pad\'e 
approximants of the flux $\overline{F}$ (Eq.~\ref{eq:secondFlux}) and the
 numerical flux in the test mass limit at 
$v=0.2, 0.25, 0.35$. The 5.5PN Taylor series, denoted by $\overline{F}_0^{11}$,
 fits the exact numerical data better than the Pad\'e approximants 
$\overline{F}^{10}_1,\overline{F}^5_6, \overline{F}^3_8, \overline{F}^2_9$.
 In the other cases, the Pad\'e approximants provide a  
better agreement (i.e. $\overline{F}^{1}_{10}$, $\overline{F}^{9}_{2}$, 
$\overline{F}^{8}_{3}$, $\overline{F}^{7}_{4}$ and $\overline{F}^{6}_{5}$, 
$\overline{F}^{4}_{7}$ and $\overline{F}^{0}_{11}$).}
\end{figure}

According to Pad\'e theory, the convergence of the Pad\'e approximant is best 
along the diagonal, but it is equally good along the off-diagonal terms if no
 pole exists in the region of interest (i.e. no zeroes appear in the 
denominator of the Pad\'e approximant.) 
For this reason, we show the error between all the possible maximal 
Pad\'e-approximated fluxes $\overline{F}^{11-n}_n$ (Eq.~\ref{eq:firstFlux}) and
the numerical flux for three values of $v$ $(=0.2$, $0.25$, $0.35)$ 
($x=0.04$, $0.06$, $0.12$) in Fig~\ref{fig:TestMass-maximalpade}. 
 The 5.5PN Taylor series, denoted by $\overline{F}_0^{11}$, fits the exact
 numerical data better than the Pad\'e approximants 
$\overline{F}^{10}_1,\overline{F}^5_6, \overline{F}^3_8, \overline{F}^2_9$.
 In the other cases, the Pad\'e approximants provide a  
better agreement (i.e. $\overline{F}^{1}_{10}$, $\overline{F}^{8}_{3}$, 
$\overline{F}^{7}_{4}$ and $\overline{F}^{6}_{5}$) for the three values of $v$.
This suggests that the Pad\'e 
approximation should only be used to suggest a fitting formula for the 
numerical data, since there is no internal self-consistency in the agreement.
The off-diagonal approximants do not 
show any regular pattern of convergence to the numerical data nor are they 
better than the Taylor series. 

%%%%%%%%%%%%%%%%%%%%%%%%%%%%%%%%%%%%%%%%%%%%%%%%%%%%%%%%%%%%%%%%
%%%%%%%%%%%%%%%%%%%%%%%%%%%%%%%%%%%%%%%%%%%%%%%%%%%%%%%%%%%%%%%%

%%%%%%%%%%%%%%%%%%%%%%%%%%%%%%%%%%%%%%%%%%%%%%%%%%%%%%%%%%%%%%%%
%%%%%%%%%%%%%%%%%%%%%%%%%%%%%%%%%%%%%%%%%%%%%%%%%%%%%%%%%%%%%%%%

%%%%%%%%%%%%%%%%%%%%%%%%%%%%%%%%%%%%%%%%%%%%%%
\subsection{Flux for the equal mass case }
%%%%%%%%%%%%%%%%%%%%%%%%%%%%%%%%%%%%%%%%%%%%%%

 For binaries of comparable mass on a quasi-circular orbit, the flux is 
known only to 3.5PN order~\cite{Blanchet2006}. 
In Ref.~\cite{Toprep} for a quasi-circular non-spinning binary, 
the numerical flux was computed by integrating 
the spin-weighted spherical harmonic components of the Weyl scalar $\Psi_4$. 
 The numerical flux data we use in this paper was provided by Harald P. 
Pfeiffer and Michael Boyle.
The estimated error in  measuring  the flux data was about 0.2\%. The velocity 
range for the simulation was from $v\sim0.26$ ($x\sim0.06$) to $v\sim0.4$ 
($x\sim0.16$). 

In the equal mass case, we cannot do an accurate convergence test early in
 the evolution as in 
Fig.~\ref{fig:TestMass-FluxConvergence} for two reasons. The first reason is
the ``junk radiation'' (noise early in the evolution from imprecise initial 
data) during the first few 
orbits. The second reason is the
 inability to accurately define the numerical flux as a function of the 
orbital frequency of the binary. The numerical normalized flux is computed 
as a function of $\omega_{22}/2$, where $\omega_{22}$ is the wave frequency 
of the $\dot{h}_{22}$ mode. Instead, in 
Table~\ref{tab:EqualMass-ConvergenceTest} we compare the convergence of the 
four flux functions  $\overline{F}_n$, $\overline{F}_m^{m+\epsilon}$, $F_n$ and
 $F_m^{m+\epsilon}$ (defined in Eqs.~\ref{eq:firstFlux}, \ref{eq:secondFlux},
 \ref{eq:thirdFlux} and \ref{eq:fourthFlux} respectively as a function of PN 
order) for $v=0.2$ 
($x=0.04$), $v_{\rm pole}=0.69$  ($x_{\rm pole}=52/109$) and $v_{\rm lso}=0.50$
($x_{\rm lso}=0.254$). We use the last stable orbit 
frequency corresponding to the 3PN Taylor series of the energy and the pole 
corresponding to $E^1_1$. The convergence does not depend on these values
although the flux values listed in Table~\ref{tab:EqualMass-ConvergenceTest} 
do depend somewhat on the values of $v_{\rm pole}$ and $v_{\rm lso}$. 
We choose a medium velocity ($v=0.2$) to make the rate of convergence clear. 
 At 3.5PN order, all four flux functions agree to 2 significant 
figures. However after seven terms, $\overline{F}_n$ converged to 3 significant
figures, $\overline{F}_m^{m+\epsilon}$ converged to 4 significant figures, 
while $F_n$ and $F_m^{m+\epsilon}$ converged to 2 significant figures. The flux
 function $\overline{F}_m^{m+\epsilon}$ converged to one additional significant
 figure over $\overline{F}_n$, however $\overline{F}_m^{m+\epsilon}$ cannot 
reliably be considered more accurate than $\overline{F}_n$ because it converges
to a slightly different value.
The Pad\'e approximants do not seem to converge to a larger number of 
significant figures than the Taylor flux function $\overline{F}_n$.

\begin{figure}
\begin{center}
\psfrag{v}[c]{\large $v$} %large
\includegraphics[scale=0.47]{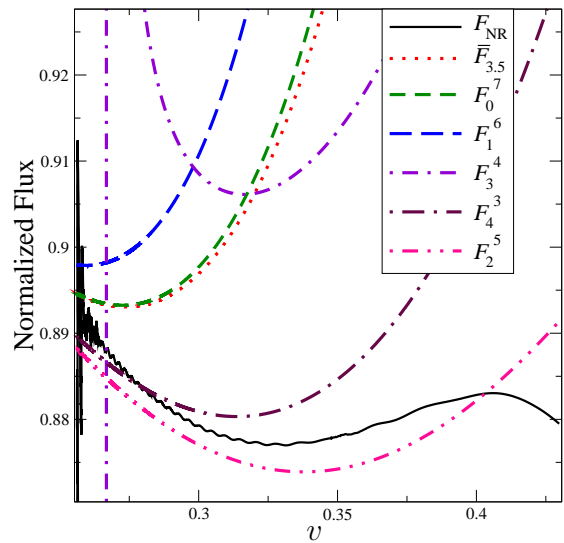}
\end{center}
\caption{\label{fig:normalizedflux} Normalized flux for an equal mass 
nonspinning binary. 
We plot the numerical flux $F_{NR}$, the 
3.5PN flux $\overline{F}_{3.5}$ and the maximal Pad\'e-approximated flux
 functions $F_4^3$, $F_3^4$, $F_2^5$, $F_1^6$ and $F_0^7$ ($\equiv F_7$).
The early noise is caused by the junk radiation.}
\end{figure}

In Fig.~\ref{fig:normalizedflux}, we plot the numerical normalized flux 
$F_{NR}$, the 3.5PN flux $\overline{F}_{3.5}$ and the maximal 
Pad\'e-approximated flux functions $F_4^3$, $F_3^4$, $F_2^5$, $F_1^6$ and 
$F_0^7$ ($\equiv F_7$). Although $\overline{F}_{3.5}$ diverges from the
 numerical flux early at $v\sim0.26$, it still fits the numerical data better
 than $F_3^4$, $F_1^6$ and $F_0^7$. The quantity $F_3^4$ has a pole and fails
 to capture the
 numerical flux behavior completely. The quantity $F_0^7$ is by definition the
 Taylor flux 
with a pole, $F_7$. This function shows that adding a pole to the Taylor
 expansion of the flux $\overline{F}_{3.5}$ degrades the fit with the numerical
 flux. Moreover, the numerical flux does not suggest the existence 
 of a pole at a large velocity ($v\sim0.69$); it starts to decrease to 0 at 
$v\sim0.4$. Adding a pole does not seem a useful idea in this case at least. On
 the other hand, $F_2^5$ and $F_4^3$ are a better fit to the numerical data 
during most of the velocity range of the 15-orbit data. The flux function 
$F_2^5$ is especially a good fit to the numerical normalized flux at high
 velocities. However, even though $F_2^5$ and $F_4^3$ are a good fit to the
 numerical flux during the last 15-orbit inspiral before merger, there is no
 guarantee that this is true at low velocities.

%%%%%%%%%%%%%%%%%%%%%%%%%%%%%%%%%%%%%%%%%%%%%%%%%%%%%%%%%%%%%%%%%%%
\begin{table}
\begin{tabular}{c|c|c|c|c}
PN order & $\overline{F}_n$  & $\overline{F}_m^{m+\epsilon}$ & $F_n$  & $F_m^{m+\epsilon}$  
\\\hline
0.0&	1.000000&	1.000000&	1.407582&       1.407582
\\
0.5&	1.000000&	1.000000&	1.000000&	1.000000
\\
1.0&	0.822381&	1.000000&	0.749987&	0.353292
\\
1.5&	{\bf 0.9}22912&	0.886577&	0.963887&	0.865262
\\
2.0&	{\bf 0.9}22745&	{\bf 0.9}05792&	0.922678&	{\bf 0.91}0047
\\
2.5&	{\bf 0.9}04387&	{\bf 0.91}0595&	0.896904&	{\bf 0.91}2033
\\
3.0&	{\bf 0.913}204&	{\bf 0.9122}61&	{\bf 0.91}6323&	{\bf 0.91}2613
\\
3.5&	{\bf 0.913}314&	{\bf 0.9122}23&	{\bf 0.91}3275&	{\bf 0.91}1492
\end{tabular}
\caption{\label{tab:EqualMass-ConvergenceTest} 
Flux convergence in the equal mass case for $v=0.2$ 
($x=0.04$), $v_{\rm pole}=0.69$ ($x_{\rm pole}=52/109$) and $v_{\rm lso}=0.50$ 
($x_{\rm lso}=0.254$).
The four flux functions $\overline{F}_n$, $\overline{F}_m^{m+\epsilon}$, 
$F_n$ and $F_m^{m+\epsilon}$ are given in Eqs.~\ref{eq:firstFlux}, 
\ref{eq:secondFlux}, \ref{eq:thirdFlux} and \ref{eq:fourthFlux} respectively. 
At 3.5PN order, all four flux functions agree to 2 significant figures. 
After seven terms, $\overline{F}_n$ converges to 3 significant figures, 
$\overline{F}_m^{m+\epsilon}$ converges to 4 significant figures, while 
$F_m$ and $F_m^{m+\epsilon}$ converge to 2 significant figures.}
\end{table}
%%%%%%%%%%%%%%%%%%%%%%%%%%%%%%%%%%%%%%%%%%%%%%%%%%%%%%%%%%%%%%%%%

%###############################################
\section{Pad\'e Waveforms}
%###############################################
\label{sec:waveforms}

 The construction of the post-Newtonian waveforms requires solving the
post-Newtonian equations describing the motion of the binary and the
 generation of gravitational
waves. Substituting the orbital evolution
predicted by the equations of motion into the expressions for the
waveform would not generate waveforms accurate enough for matched
filtering in detecting gravitational waves~\cite{cutler_etal93}. 
To compute the waveform at 3PN order, it is necessary to solve the equations 
of motion at 5.5PN order, because the radiation reaction contributes to the 
equations
of motion starting at 2.5PN order.
However, for a non-spinning binary of equal mass and 
on a circular orbit, accurate waveforms at 3PN order can be constructed under 
two further 
assumptions. The first assumption is that the binary follows a slow adiabatic 
inspiral. The second assumption is that of 
 energy balance between the orbital binding energy 
and the energy emitted by the gravitational waves,
where the energy balance equation is defined as
\begin{equation}
\label{eq:EnergyBalance}
\frac{dE}{dt} = - {F}.
\end{equation}
%While this is extremely plausible, it has only been confirmed through
%1.5 PN order\cite{Blanchet97}.

%%%%%%%%%%%%%%%%%%%%%%%%%%%%%%%%%%%%%%%%%%%%%%%%%%%%%%%%%%%%%%%%
%%%%%%%%%%%%%%%%%%%%%%%%%%%%%%%%%%%%%%%%%%%%%%%%%%%%%%%%%%%%%%%%

The procedure of constructing the standard Pad\'e waveforms~\cite{Damour98} 
is similar to one used to 
 construct the TaylorT1 waveforms in Ref.~\cite{Damour98,Boyle2007}.
The main difference is the use of Pad\'e approximants of 
the energy and flux to compute the orbital phase, as described in 
Secs.~\ref{sec:energy} and \ref{sec:flux}, instead of their Taylor 
expansions. 
The orbital phase used in the Pad\'e waveforms is obtained by numerically 
integrating 
\begin{equation}
\label{eq:padefluxenergy}
\frac{d\Omega}{dt} = \frac{32}{5} \nu^2 v^{10} \frac{F^m_n}{dE^k_l/d\Omega}\,.
\end{equation}
The fraction on the right side of
Eq.~(\ref{eq:padefluxenergy}) is retained as a ratio of the Pad\'e approximants
of the post-Newtonian
expansions, and is not expanded further before numerical integration.
The waveform is produced by substituting the orbital phase into 
the spherical harmonic mode $h_{22}$ of the post-Newtonian waveform, 
which is known up to 3PN order~\cite{Kidder07a,ABIQ,ABIQ-erratum,BFIS}.

Given the expressions for the Pad\'e-approximated energy and flux in 
Sections~\ref{sec:energy} and \ref{sec:flux}, 
and the Taylor series of the waveform amplitude~\cite{Kidder07a,ABIQ,ABIQ-erratum,BFIS}, 
there is still a set of choices that must be made in order
to produce a Pad\'e-approximated waveform that can be compared to our numerical
waveform. These include
\begin{enumerate}
\item The Pad\'e approximant of the orbital energy, $E^k_l$.
\item The flux function and its Pad\'e approximant $F^n_m$.
\item The velocity of the pole and the last stable orbit, $v_{\rm pole}$ and 
$v_{\rm lso}$.
\item The PN order through which terms in the waveform amplitude
  are kept.
\end{enumerate}

\subsection{Procedure}

We consider numerical gravitational waves extracted with the Newman-Penrose 
scalar $\Psi_4$, using the same procedure as
in~\cite{Pfeiffer-Brown-etal:2007}. To minimize gauge effects, we compare
 its $(2,2)$ component extrapolated to 
infinite extraction radius according to Ref.~\cite{Boyle2007}.
The extracted waveform is split into real phase $\phi$ and real amplitude $A$,
defined by Ref.~\cite{Boyle2007} as 
\begin{equation}\label{eq:A-phi-definition}
\Psi^{22}_4(r,t) = A(r,t)e^{-i\phi(r,t)}.
\end{equation}
The gravitational-wave frequency is given by
\begin{equation}\label{eq:omega-definition}
\omega=\frac{d\phi}{dt}.
\end{equation}
The spherical harmonic component (2,2) of $\Psi_4$ is then compared to the 
 numerically twice-differentiated post-Newtonian expression of $h_{22}$, 
$A_{22}$, as in Ref.\cite{Boyle2007}. 
Following~\cite{Baker2006d,Hannam2007,Boyle2007}, 
the matching procedure needed to set the arbitrary time offset $t_0$ and
 the arbitrary phase offset $\phi_0$ is done by demanding that the PN and NR 
gravitational wave phase and gravitational wave frequency agree at some 
fiducial frequency $\omega_M$.

\subsection{Results}

%%%%%%%%%%%%%%%%%%%%%%%%%%%%%%%%%%%%%%%%%%%%%%%%%%%%%%%%%%%%%%%%
\begin{figure}
%\label{fig:padewaveform}
%\includegraphics[scale=0.47]{Pade-FPxxExx2}
\includegraphics[scale=0.47]{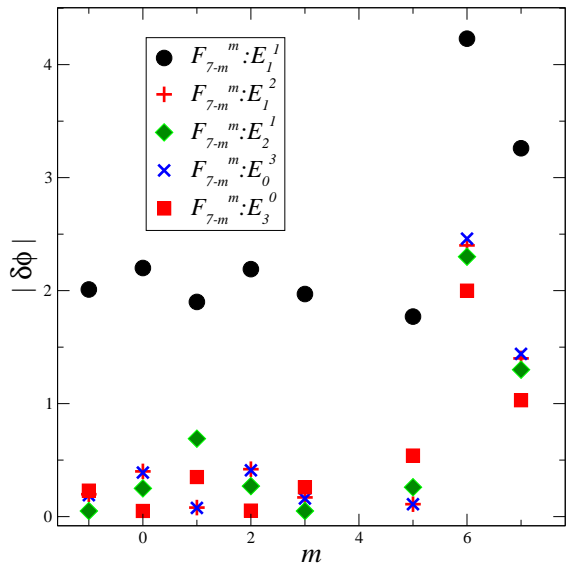}
\caption{\label{fig:padewaveform} 
Phase difference between the 3 and 3.5 PN Pad\'e approximated and 
numerical waveforms matched at the wave frequency $M\omega=0.1$. 
We use the Pad\'e-approximated flux $F^m_{7-m}$ (Eq.~\ref{eq:fourthFlux}) and
 energy $E^k_l$. We include in the figure the waveforms using the 
Pad\'e-approximated flux $F^3_3$ using $m=-1$.  There is no entry
 for $m=4$ since the Pad\'e-approximated flux $F^4_3$ has a pole in the 
frequency range of the simulation.}
\end{figure}
%%%%%%%%%%%%%%%%%%%%%%%%%%%%%%%%%%%%%%%%%%%%%%%%%%%%%%%%%%%%%%%%

In this section, we compare the numerical waveform to the Pad\'e waveforms 
corresponding to the 3.5 PN order of energy and flux using the 3PN Taylor 
series of the post-Newtonian amplitude $A_{22}$. 
The energy and flux functions used are those suggested by Ref.~\cite{Damour98}.
We do not generate all possible waveforms using different 
Pad\'e approximants of the energy or the flux at low PN orders, since all 
these resummed series showed no improvement in the convergence rate.  

As introduced in Sec.~\ref{sec:energy}, we use 
the Pad\'e-approximated energy $E^3_0$, 
$E^2_1$, $E^1_2$ and $E^0_3$ corresponding to the £PN Taylor series of 
the energy, and the Pad\'e-approximated energy, $E^1_1$, 
corresponding to its 2PN Taylor expansion. 
For the flux,  
the diagonal Pad\'e approximant $F^3_3$ is used in addition
to all possible Pad\'e approximants of flux at 3.5PN order, $F^m_{7-m}$, 
where $0\le m \le 7$ as described in Sec.~\ref{sec:flux}.
  
The Pad\'e-approximated flux has two parameters, $v_{\rm lso}$ and 
$v_{\rm pole}$ as discussed in Sec.~\ref{sec:flux}. The value $v_{\rm pole}=52/109$
 is used. We also tested varying the pole location, but found that we could not
 improve the agreement significantly. 

 From Table~\ref{tab:polelso}, any value of the velocity of the last stable 
orbit could be used. We use the 3PN value $v_{\rm lso}=0.254$ and also use
 $v_{\rm lso}=0.199$. The latter is used when the Pad\'e approximant $E^1_1$ is
employed in the construction of the waveform. In the remaining cases, we use
 $v_{\rm lso}=0.254$ since it is quite close to the estimates from other Pad\'e
 approximants of the energy. The effect of changing the value of 
$v_{\rm lso}$ is not significant compared to changing the order of the Pad\'e 
approximant for the energy or the flux.

To do the comparison, we match the Pad\'e-approximated and numerical 
waveforms at the wave frequency $M\omega=0.1$. Then we measure the maximum
 phase difference between the numerical waveform and each of these Pad\'e 
waveforms during the inspiral when the numerical wave frequency is between   
$M\omega=0.035$ and $M\omega=0.1$ (as in the upper panel of 
Fig.~\ref{fig:phasedifference}).  Our results are summarized in Fig.~\ref{fig:padewaveform},
which
 shows the phase differences for each of the Pad\'e approximants of energy 
$E^k_l$ and flux $F^m_{7-m}$. On the same Figure, we include phase 
differences for the waveforms generated using the 
Pad\'e-approximated flux $F^3_3$ under the $m=-1$ entry. 

 When $E_1^1$ is used, the phase error ranges between $2$ and $5$ 
radians as $m$ increases from $-1$ to $7$. Using all the possible Pad\'e 
approximants of the 3PN energy, the estimated phase difference ranges from 
$0.05$ to $2.5$ radians. Using the Taylor series with a pole ($m=7$) 
resulted in a large phase difference ranging between 1 and 1.5 radians.
The diagonal Pad\'e term $F^3_4$ of the flux generates similar phase 
differences, ranging from $0.06$ to $0.2$ radians as the Pad\'e order of the 
energy changes. 

The Pad\'e-approximated waveforms do not fit the numerical data better
 than the waveforms using the Taylor expansion of the flux.
Although the Pad\'e-waveforms along the diagonal have a phase difference less 
than 0.25 radians, none of these waveforms fits the numerical waveforms
 better than  TaylorT4 at 3.5PN order as shown in Ref.~\cite{Boyle2007}. 
Moreover, the dependence of the phase difference on the Pad\'e order suggests 
that there is no reason why it should help in estimating the parameters better 
in data analysis. This is as expected from the poor  
convergence of the the Pad\'e approximant of the flux discussed in 
Sec.~\ref{sec:flux}.

The Pad\'e resummation techniques were also tested on the Taylor series for the
amplitude, and they showed no improvement in the convergence of the series.  
In adition, none of the tests that were performed on the Pad\'e resummed Taylor
 series of the T2 and T3 waveforms showed a faster convergence rate. 
In fact, there is no improvement in convergence for any Taylor series in the PN
approximation that we have investigated.

%###############################################
\section{Simple EOB Model}
%###############################################
\label{sec:simpleeob}

We have described  the failure of the Pad\'e resummation techniques to 
accelerate the convergence of any PN Taylor series, the absence of any 
signature of a pole in the flux in the equal mass case, and the erratic pattern
 of agreement between the Pad\'e waveforms and the numerical waveform. It seems
 one might as well simply use the Taylor series 
at all steps of computing waveforms. Also it does not seem that the parameters 
$v_{\rm pole}$ and $v_{\rm lso}$ are useful. In this section, we show how to
 get good agreement with the numerical waveform by using a simple EOB
 model. The only parameter we introduce and fit for is an unknown 4PN 
contribution to the flux.

\subsection{EOB waveforms}

The EOB formalism~\cite{Damour2008} is a non-perturbative analytic approach 
that handles 
the relative dynamics of two relativistic bodies. This approach of resumming 
the PN theory is expected to extend the validity of the PN results into the 
strong-field limit. The procedure for generating an EOB waveform follows 
closely the steps in Sec.~\ref{sec:waveforms}. Instead of using the 
energy balance equation, we compute the orbital phase  by numerically 
integrating Hamilton's equations. The EOB waveform is generated by substituting
 the orbital phase into the waveform amplitude $A_{22}$ at 3PN order.
 The two fundamental ingredients that allow computing the
orbital phase are the real Hamiltonian $\hat{H}$ and the the radiation reaction
$\cal F_\phi$.

\subsection{Hamilton's equations}

In terms of the canonical position variables $r$ and $\phi$ and their conjugate
 canonical momenta $p_r$ and $p_\phi$, where $r$ is the relative 
separation and $\phi$ is the orbital phase, the real dynamical Hamiltonian is 
defined as~\cite{Damour:2007cb}:
%\begin{widetext}
\begin{equation}
\hat{H}=\frac{1}{\nu} \sqrt{1+ 2 \nu \left(H_{EOB}-1\right)}\,,
\end{equation}
%\end{widetext}
where
\begin{equation}
H_{EOB}=\sqrt{A\Big( 1 + \frac{p_{\phi}^2}{r^2} + \frac{p_r^2}{B} +2\nu(4-3\nu) \frac{p_{r}^4}{r^2}  \Big) }\,,
\end{equation}
and where the radial potential $A$ function is defined as the series
\begin{equation}
A= 1 -\frac{2}{r}+\frac{2 \nu}{r^3}+\Big( \frac{94}{3}-\frac{41}{32} \pi^2\Big) \frac{\nu}{r^4}\,.
\end{equation}
The Taylor series of the $A$ function is replaced by its Pad\'e approximant 
$A^1_3$. Here the Pad\'e approximant 
is not used to accelerate the convergence of the Taylor expansion 
of $A$. Instead, it leads to the existence of a last stable orbit (see 
Ref.~\cite{Damour2007a} and references therein). Otherwise, the EOB 
Hamiltonian  
is non-physical for the last few orbits; the orbital frequency stays nearly 
constant for several orbits before merger. For the $B$ function, the 
Taylor expansion suffices:  
\begin{equation}
B= \frac{1}{A} \Big[ 1 -\frac{6\nu}{r^2}+2(3\nu-26)\frac{ \nu}{r^3}\Big]\,.
\end{equation}
Then Hamilton's equations of motion are given 
in the quasi-circular case by
\begin{eqnarray}
\partial_t r &=& \partial_{p_r} \hat{H} \,,
\\
\partial_t \phi &=& \partial_{p_\phi} \hat{H}\,, 
\\
\partial_t p_r &=& -\partial_{r} \hat{H} \,,
\\
\label{eq:hamilton4}
\partial_t p_\phi &=& -\cal F_\phi\,,
\end{eqnarray}
where $\cal F_\phi$ is the radiation reaction in the $\phi$ direction 
representing the nonconservative part of the dynamics. 
In Eq.~\ref{eq:hamilton4}, $\partial_{\phi} \hat{H}=0$ since  
$\hat{H}$ is independent of $\phi$.
The radiation reaction is deduced from the post-Newtonian flux as in  
Refs.~\cite{Iyer93,Iyer95,Buonanno00}
\begin{equation}
{\cal F}_\phi = \frac{F+F_8 v^8}{\nu v^3}\, . 
\end{equation}
In this equation, we have introduced an unknown 4PN flux term, $F_8$, the only 
parameter that we fit for in this EOB model. 

\subsection{Initial conditions}

To integrate Hamilton's equations,
we need appropriate initial conditions for a quasi-circular orbit.  
Refs.~\cite{Buonanno00,Damour03,Buonanno06} indicate how to
define some ``post-adiabatic'' initial conditions. However, these initial 
conditions do not generate an orbit with as low an eccentricity as the 
numerical simulation, roughly $5\times 10^{-5}$. 
At a given radius $r$, starting from the 
post-adiabatic initial conditions of $p_r$ and $p_\phi$, 
we therefore reduce the eccentricity iteratively in two
steps. The first step includes evolving Hamilton's equations in the 
conservative regime (${\cal F}=0$) and iteratively changing the value of 
$p_\phi$ until the eccentricity measured from the evolution of the orbital
 separation is of
 the order $10^{-9}$. The second step is based on evolving the nonconservative 
Hamilton's equations with the 4PN flux and iteratively changing the $p_r$ 
momentum 
until the eccentricity is again of the order $10^{-5}$. This circularization
 procedure is repeated as we iterate $F_8$ to maximize the agreement 
between the waveforms. 

\subsection{Best Fit of $F_8$}

To find the best fit for $F_8$, we iteratively solve for the minimum in the 
phase 
difference between the numerical and EOB waveforms. The waveforms are 
matched 
as in Sec.~\ref{sec:waveforms} at the wave frequency $m\omega=0.1$, and the 
phase difference is defined as the maximal phase difference during the inspiral
phase up to the wave frequency $m\omega=0.1$. We find a best fit 
value $F_8=-333.75$ corresponding to the initial conditions $r=17$, $\phi=0$,
$p_r=-0.0008$, $p_\phi=4.53235$. A change of 1\% in $F_8$ changes the maximal 
phase difference from less than 0.002 radians to about 0.01 radians.
Note that without adding the fitting parameter $F_8$, the phase difference is
 about 1.7 radians during the 15-orbit inspiral.

\subsection{Results}

%%%%%%%%%%%%%%%%%%%%%%%%%%%%%%%%%%%%%%%%%%%%%%%%%%%%%%%%%%%%%%%%
\begin{figure}
\includegraphics[scale=0.47]{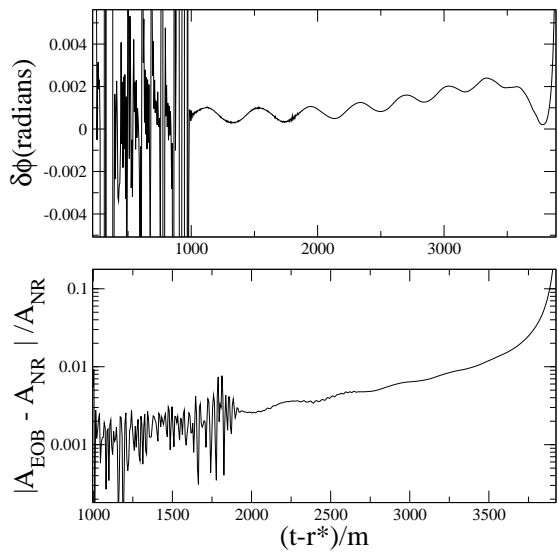}
\caption{\label{fig:phasedifference} Phase and amplitude differences between 
the EOB waveform and  the numerical waveform. After fitting for 
the best value of $F_8$, the phase difference is less than $0.002$. The early
noise is due to junk radiation at the early stage of the numerical simulation. 
$r^*$ is the tortoise coordinate defined in~\cite{Boyle2007}.}
\end{figure}
%%%%%%%%%%%%%%%%%%%%%%%%%%%%%%%%%%%%%%%%%%%%%%%%%%%%%%%%%%%%%%%%

In the upper panel of Fig.~\ref{fig:phasedifference}, we plot the phase 
difference 
between the numerical waveform and the EOB waveform computed using the 3PN 
Taylor series of the amplitude $A_{22}$. The phase difference is less than
$0.002$ radians after maximizing the agreement between the waveforms in the
region where $m\omega\leq0.1$. The early noise is due to junk radiation at
the early stage of the numerical simulation as described in Sec II C of 
Ref.~\cite{Boyle2007}. The phase uncertainty in the simulation was estimated to
be $0.05$ radians; See Table III in \cite{Boyle2007}. 

In the lower panel of Fig.~\ref{fig:phasedifference}, we plot the relative 
difference between the amplitude of the numerical waveform and the EOB 
waveform. The EOB waveform amplitude does not fit the numerical waveform 
amplitude as well as the wave phase does. This is expected because the waveform
amplitude is known to 3PN order only, and no free parameter in the amplitude 
was fitted for. The agreement between the amplitude of this EOB model and the 
numerical waveform is similar to the agreement between the amplitude of 
TaylorT4 3.5/3.0 and the numerical waveform in Fig. 21 in \cite{Boyle2007}.

This EOB model is a modification of the ET EOB model of 
Ref.~\cite{Buonanno2004}. It fits the numerical phase very well without using
 the Pad\'e resummation techniques nor a pole in the flux.

Even though we have found very good agreement between the waveforms, these 
results only suggest that the EOB model is a very good fitting model.
Moreover, having fit a particular waveform, there is no guarantee the model
will have predictive power for a more general case. 

%\hspace{1cm}

%#######################
\section{Conclusions}
%#######################
\label{sec:conclusions}

Convergence tests show that none of the Taylor series in the PN approximation, 
such as the energy or the flux, could be replaced by a Pad\'e approximant 
that converges faster. Other attempts we tried to accelerate the convergence 
of these series also failed, as for example using the Levin method to 
accelerate convergence~\cite{nr3}. As a result, more reliable waveforms could 
not be constructed using a Pad\'e resummation scheme. 
Moreover, the Pad\'e waveforms also do not 
fit numerical simulation data better than the Taylor waveforms. 
Thus, they do not seem to be better than the Taylor waveforms in building 
templates for waveforms.
This conclusion is independent of the Pad\'e approximants used to test the 
convergence. Taking for example the sub-diagonal Pad\'e approximant does 
not show any improvement in the convergence rate. In addition, this conclusion
is independent of the numerical data we used. We can simply take the highest 
PN order of the Taylor series or the Pad\'e approximant and use it as the 
``exact'' value of the function to test the convergence at low frequency.   

Based on the dependence of the flux on the velocity 
in the equal mass case, we do not find it helpful to add a pole to the flux. 
Therefore, we recommend using Taylor series instead of
the Pad\'e approximant to generate waveforms both in the time and frequency 
domains. The simple EOB model used in this paper agrees with the numerical 
data very well; the phase difference during the inspiral is much less then
the estimated phase uncertainty in the numerical data.
This model does not use Pad\'e 
approximants or poles except in one place to enforce a last stable orbit. 
Since Pad\'e approximation does not accelerate the convergence of any PN 
Taylor series, there is no reason why it should estimate parameters better 
in data analysis of waveforms.

\begin{acknowledgments} 
It is a pleasure to acknowledge useful discussions with Emanuele Berti, 
Michael Boyle, Alessandra Buonanno, Lee Lindblom, Harald P. Pfeiffer, Yi Pan,
 Mark A. Scheel and Nicol\'as Yunes. We thank Jihad Touma for helpful 
discussions about Pad\'e approximants, Harald P. Pfeiffer and Michael 
Boyle for providing the numerical data of the flux in the equal mass case, 
 and Eric Poisson for providing the 
numerical data of the flux in the test mass limit. This work was supported in
 part by grants from the Sherman Fairchild Foundation to Cornell; by NSF grants
 PHY-0652952, DMS-0553677, PHY-0652929, and NASA grant NNG05GG51G at Cornell. 
\end{acknowledgments} 

%%%%%%%%%%%%%%%%%%%%%%%%%%%%%%%%%%%%%%%%%%%%%%%%%%%%%%%%%%%%%%%%%%%%%%%%%%%%%%%
%\section*{References}
%%%%%%%%%%%%%%%%%%%%%%%%%%%%%%%%%%%%%%%%%%%%%%%%%%%%%%%%%%%%%%%%%%%%%%%%%%%%%%%
\bibliography{References/References}
%\begin{thebibliography}{99}
%\end{thebibliography}

\end{document}